\documentclass[preprint, amsmath,amssymb, aps]{revtex4-2}
\usepackage[utf8]{inputenc}
\usepackage{graphicx}
\usepackage{amsmath}
\usepackage{enumerate}
\usepackage[T1]{fontenc}
\usepackage{tipa}
\usepackage{xcolor}
\newcommand{\avg}[1]{\left \langle #1  \right\rangle}
\usepackage[version=4]{mhchem}
\usepackage{lineno}

\definecolor{britishgreen}{rgb}{0.0, 0.26, 0.15}

\usepackage{ulem}

\newcommand{\pt}{P(T)}
\newcommand{\pn}{P(N)}
\newcommand{\pr}{P(\Delta r,X)}
\newcommand{\T}{$T$ }

\begin{document}

\preprint{APS/123-QED}

\title{Modeling ball possession dynamics in the game of Football}

\author{A. Chacoma}
\email{achacoma@famaf.unc.edu.ar}
\affiliation{Instituto de F\'isica Enrique Gaviola (IFEG-CONICET).}

\author{N. Almeira}
\affiliation{Instituto de F\'isica Enrique Gaviola (IFEG-CONICET).}
\affiliation{Facultad de Matem\'atica, Astronom\'ia, 
F\'isica y Computaci\'on, Universidad Nacional de C\'ordoba.}

\author{J.I. Perotti}
\affiliation{Instituto de F\'isica Enrique Gaviola (IFEG-CONICET).}

\author{O.V. Billoni}
\affiliation{Instituto de F\'isica Enrique Gaviola (IFEG-CONICET).}
\affiliation{Facultad de Matem\'atica, Astronom\'ia, 
F\'isica y Computaci\'on, Universidad Nacional de C\'ordoba.}

\begin{abstract}

In this contribution, we study the interaction dynamics in the game of football--soccer in the context of ball possession intervals.
To do so, we analyze a novel database, comprising one season of the five major football leagues of Europe.
Using this input, we developed a stochastic model based on three agents: two {\it teammates} and one {\it defender}. 
Despite its simplicity, the model is able to capture, in good approximation,
the statistical behavior of possession times, passes lengths, and number of passes performed.
In the last part, we show that the model's dynamics can be mapped into a Wiener process with drift and an absorbing barrier. 
\end{abstract}

\maketitle

\section{Introduction}

The statistical analysis of competing games based on data gathered from professional competitions is currently a growing area of research~\cite{petersen2020renormalizing,neiman2011reinforcement, mukherjee2019prior, merritt2013environmental, mandic2019trends,perotti2013innovation,schaigorodsky2014memory,almeira2017structure}.  
In the case of team sports games, these studies have a potentially high impact.
It is boosted by commercial interests but also by its intrinsic  complexity that caught the attention of basic 
research~\cite{petersen2020renormalizing,neiman2011reinforcement, mukherjee2019prior, merritt2013environmental,  mandic2019trends}.
In the context of team sports games, the emergence of complex behaviour is often observed. It arises from the interplay dynamics of a process governed by well--defined spatiotemporal scales. 
It is well known that these scales are important for both individual interactions among athletes and collective strategies~\cite{lebed2013complexity}. 

Particularly interesting is the game of football, where data analytics have been successfully tackled in recent years~\cite{lopes2019entropy, rossi2018effective, bransen2019measuring}.
For instance, in the field of complex systems, J. Buldú et al. used network theory to analyze the Guardiola's {\it F.C.~Barcelona} performance ~\cite{buldu2019defining}. 
In that work, they consider a team as an organized social system where players are nodes linked during the game through coordination interactions.

Despite these recent contributions, football analytics seems to be relegated as compared to other major team sports, like basketball or baseball. 
That is why football’s team management and strategy is far from being recognized as analytics--driven.
The specific problem with football is concerned with data collection. 
Usually, the collection of data upon ball--based sports competitions is focused on what is happening in the neighborhood of the ball (on--ball actions). 
Nonetheless, in football games, an important part of the dynamics is developed far from the ball (off--ball dynamics), and this information is required to analyze the performance of football teams~\cite{casal2017possession}.
Consequently, in the game of football, on--ball actions might provide less insight for strategy and player evaluation than off--ball dynamics.

In this context, a possible solution is to improve the data gathering, a possibility often limited by a lack of resources.
From an alternative perspective, we aim to define a framework based on the use of state of the art statistical tools and modeling techniques, that allow us the characterization of the global dynamics by studying the local information provided by the data.

Based on these ideas, and on previous studies~\cite{yamamoto2018examination, hunter2018modeling, cakmak2018computational}, 
in the present contribution, we have surveyed, collected and analyzed information from a novel database~\cite{pappalardo2019public}
to propose an innovative agent--based football model.
We emphasize, our goal is not to model the full complexity dynamics of a football game, but to model the dynamics of ball possession intervals, defined as the consecutive series of actions carried out by a team. We focus on study the interactions in the frame of both on-ball and off-ball actions, considered as the main feature to understand the team's collective performances \cite{gama2016networks, gudmundsson2017spatio}.

This paper is organized in three parts: Material and Methods, Results and Discussion. In section Material and Methods, we firstly introduce the database. In particular, we describe the dataset {\ttfamily Events}, as well as other information regarding relevant fields.
Secondly, we discuss some interesting statistical patterns that we found in this dataset, and we use to propose the model's components.
Thirdly, we give a formal definition of the model and discuss in detail the key elements, the assumptions, and the dynamical parameters.
Lastly, we present a method to systematically search for a suitable set of parameters for the model. 
The section Results is divided into two parts. Firstly we evaluate the results of the model.
To do so, we focus on analyzing three statistical observables (i) the distribution of possession time, (ii) the distribution of the distance traveled by the ball in passes (hereafter referred to as the passes length), and (iii) the distribution of the number of passes.
The idea is to assess the model's performance by comparing its outcomes with the data. 
Secondly, we place our model in a theoretical framework. This allows, under certain approximations, an interpretation of the emergent spatiotemporal dynamics of the model.
Finally, our results are discussed in the last section.


\section{Material and Methods}
\label{se:metodos}

\subsection{The dataset}
\label{se:material}

In 2019, L. Pappalardo {\it et al.} published one of the largest football--soccer database ever released \cite{pappalardo2019public}.
Within the information provided in this astounding work, 
the dataset {\ttfamily Events} contains a gathering of all the spatiotemporal events recorded from each game in the season $2017$-$2018$ of the following five professional football leagues in Europe: Spain, Italy, England, Germany, and France. 
A typical entry in this dataset bears information on,

\begin{itemize}
    \item {\it Type of event.} 
    Namely, pass, duels, free kicks, fouls, etc, 
    subdivided into other useful subcategories. This field allows us to evaluate in detail the correlation between particular actions and the consequences in the dynamics.   
    
    \item {\it Spatiotemporal data.}
    Each event is tagged with temporal information, referred to the match period, and to its duration in seconds.
    Spatial information, likewise, is referred to the stadiums' dimensions as a percentage of the field length from the view of the attacking team.
    
    \item {\it Unique identifications.}
    Each event in the dataset is linked to an individual player in a particular team. 
    This allows us to accurately determine the ball position intervals, and moreover to perform a statistical analysis of the players involved.  
\end{itemize}

In light of this information, we define a ball position interval (BPI) as the set of consecutive events generated by the same team.
We gathered $3071395$ events and $625195$ BPIs from the dataset, totalizing $1826$ games, involving $98$ teams, 
and with the participation of $2569$ different players. 
Since we aim to study a dynamical evolution, only BPIs with two or more events were collected. 
On the other hand, since different games often occur in stadiums of varying sizes, to compare distances we normalized all the measured distances in a game to the average distance calculated using the whole set of measures in that game.

\subsection{Statistical patterns}
\label{se:insights}
The idea of this section is to present the statistical patterns that we have used to propose the main components of our football model. 
Firstly, in Fig.~\ref{stats} panel A, we plot the frequency of events by type (blue bars) and also the frequency of events that trigger a ball possession change (BPC) (see red bars). 
By looking at the blue bars, we can see that the most common event is the {\it "Pass"}, with $1.56$ million entries.
Notice that passes, almost duplicate the second most frequent type of event, {\it "Duels"}. Which at the same time is the most frequent event triggering the possession changes  (see red bars).
Moreover, by comparing the two bars on {\it "Duels"}, we can see that $\approx 75\%$ of the duels produce possession changes, showing that this type of event is very effective to end BPIs. 

Secondly, in Fig.~\ref{stats} panel B, the main plot shows the number of different players involved per BPI. 
As can be seen, the most common case is {\it two players}, with $0.27$ million observations, duplicating the {\it three players} case, the second most commonly observed. 
The inset shows the number of different types of events per BPI. 
With $0.4$ million of cases recorded, we can see the case of two types of events is the most common. 
Notice, the data seems to show statistical regularities.
Despite the doubtless complexity of the game, there are features that dominate over others.

In the following section, we use these observations to propose the main components of a minimalist dynamical model.

\subsection{The model}
\label{se:rules}

We aim to build a model that draws the main features of football game dynamics during ball possession intervals.
The idea is to propose a system both simple and minimalist, but also effective in capturing global emergents of the dynamics.
To do so, we used the empirical observations made in the previous section.

Let us think in a system with three agents ({\it the players}), two in the same team having possession of the ball ({\it the teammates}), and one in the other ({\it the defender}).
The players in this system can move in two dimensions, and the teammates can perform passes to each other.
In this simulated game, the system evolves until the defender reaches the player with the ball and, emulating a {\it Duel}, it ends the BPI. 
Bearing these ideas in mind, in the following we propose the rules that govern the agents' motion, and consequently define the model's dynamics.

\begin{figure*}[t!]
\centering
\includegraphics[width=1\textwidth]{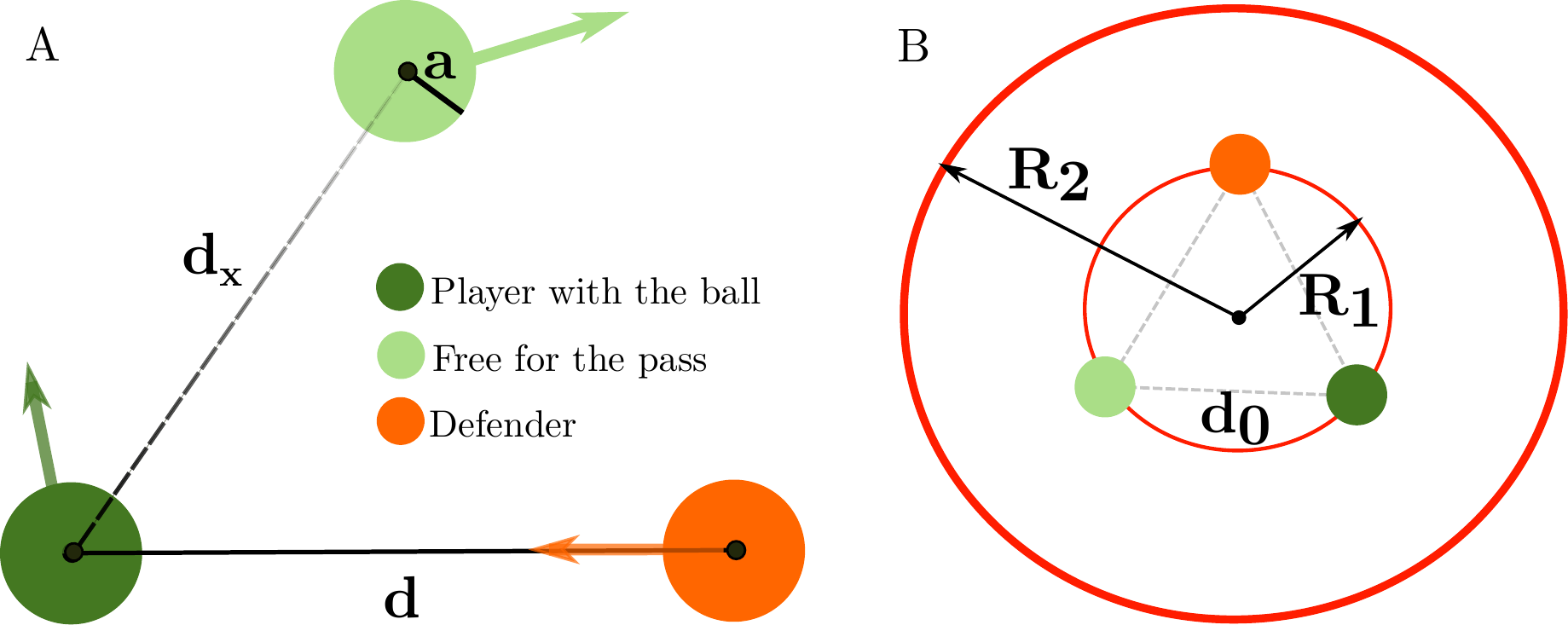}
\caption{Scheme summarizing the main parameters of the model (not to scale). 
Green circles represent the {\it teammates}, and orange circle {\it the defender}.
(A) We emphasize on the parameter : (i) $d$, the distance between the player with the ball and the defender, (ii) $d_x$, the distance between the player with the ball and the free player, and (iii) $a$, the action radius.
(B) The circles placed at distance $R_1$ from the origin, represent the initial condition in the dynamics. Distance $d_0$, is the initial distance between the three agents. 
Radius $R_2$ delimits the agents' moving area. 
}
\label{diagram}
\end{figure*}

Let $\Vec{r_i}(t)$ be a $2D$ position vector for an agent $i$ ($i=1,2,3$) at time $t$.
Considering discrete time steps $\Delta t=1$, at $t+1$ the agents will move as $\Vec{r_i}(t+1)=\Vec{r_i}(t)+ \Vec{\delta r_i}(t)$.
In our model, we propose 
$\Vec{\delta r_i}(t) = (R \cos \Theta, R \sin \Theta)$,
where $R$ and $\Theta$ are two variables taken as follows,

\begin{enumerate}

    \item {\it The displacement $R$}
    
        The three agents randomly draw a displacement from an exponential distribution  $P_a(r)=\frac{1}{a}e^{-r/a}$,
        where $a$, the scale of the distribution, is the agent's action radius  (see Fig.~\ref{diagram} A), i.e. the surroundings that each player controls.    
   
    \item {\it The direction $\Theta$}
        \begin{enumerate}
            \item {\it For the teammates}. The agents randomly draw an angle in $[0, 2\pi)$ from a uniform distribution.
            
            \item {\it For the defender}. This agent takes the direction of the action line between itself and the agent with the ball.
        \end{enumerate}
   
\end{enumerate}

Then, according to the roles in the game, the players decide to accept the changes proposed as follows, 

\begin{enumerate}
    \setcounter{enumi}{2}
    \item  {\it The player with the ball} evaluates if the proposed displacement moves it away from the defender. If it does, the player changes the position; otherwise, it remains the current position.
    
    \item {\it The free player} and {\it the defender} always accept the change.

\end{enumerate}
    
As we mentioned before, in this model we consider the possibility that the teammates perform passes to each other.
This decision is made as follows,

\begin{enumerate}
    \setcounter{enumi}{4}

    \item If the defender's action radius does not intercept the imaginary line joining the teammates, then the player with the ball plays a pass to the other teammate with probability $p$.

\end{enumerate}

Since in real football games the player's movements are confined, for instance, by the field limits, in the model we introduce two boundary parameters: The inner and external radius, $R_1$ and $R_2$, respectively, (see Fig.~\ref{diagram} B).

\begin{enumerate}
    \setcounter{enumi}{5}
    \item The inner radius $R_1$ is used to set the initial conditions. At $t=0$, each one of the three agents is put at a distance $R_1$ 
    from the center of the field, spaced with an angular separation of $120$ degrees (maximum possible distance between each other).
    
    \item The external radius $R_2$ defines the size of the field. It sets the edge of the simulation. 
    If an agent proposes a new position $\Vec{x}(t+1)$, such that $||\Vec{x}(t+1)||\geq R_2$, then, the change is forbidden, and the agent keeps its current position -- note this overrules the decision taken from $(3)$ and $(4)$. 
\end{enumerate}

Lastly, a single realization of the model in the frame of the rules proposed above ends when,

\begin{enumerate}
    \setcounter{enumi}{7}
    \item The defender invades the agent with the ball's action radius. That is, when the distance $d$ between the player with the ball and defender satisfies $d<a$.
\end{enumerate}

\begin{figure*}[t!]
\centering
\includegraphics[width=1\textwidth]{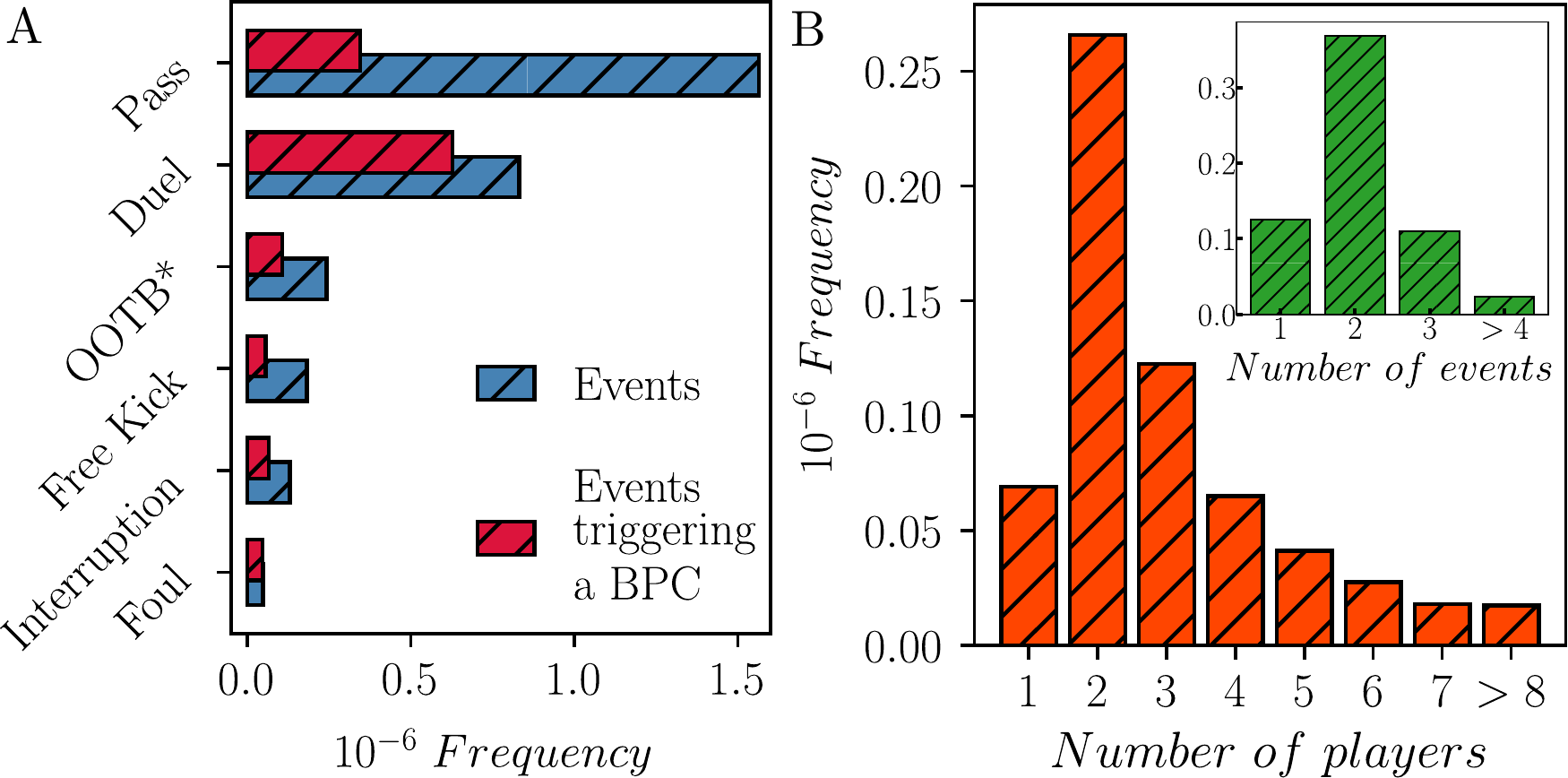}
\caption{
Relevant statistical patterns gathered from the dataset {\ttfamily Events} in ref. \cite{pappalardo2019public}. 
(A) Frequency by type of event. Blue bars, from the set of all the events. Red bars, only the events triggering a ball possession change (BPC). 
(B) The main plot shows the number of different players involved in a ball possession interval (BPI). The inset shows the number of different types of events in a BPI.
 $(*)$ The acronym OOTB, stands for {\it Others on the ball}. 
}
\label{stats}
\end{figure*}

Let us justify the election of the rules and the different elements of the model.
Firstly, it is well--known that football exhibits a complex dynamics.
Fig.\ref{stats} (A) shows that many events are possible in the context of a BPI. However, we can see that the events {\it "Pass"} and {\it "Duels"} domain in the frequency of the common events, and events triggering a BPC, respectively. 
Therefore, a reasonable simplification is to propose a model with only two possible events. This also agrees with the data shown in the inset of Fig.~\ref{stats} (B), regarding the number of different types of events observed during BPI.

Secondly, considering only three players for a football model could be seen as an oversimplification.
However, as we show in the main plot of Fig.~\ref{stats} (B), the number of players by BPI is in most of the cases two. 
Therefore, a system with two teammates and a single defender triggering the BPCs is, presumably, a good approximation; ultimately, to be judged by the model's predictions on the observed statistics.

Thirdly, let us discuss the players' movement rules. In item (1) (see listing above), we propose the agents raw the displacements from an exponential distribution, with an action radius $a$ as the scale.
The idea behind this is to set a memoryless distribution, in the light that the players' displacements are commonly related to both evasion and distraction maneuvers, which are more effective without a clear motion pattern~\cite{bloomfield2007physical}. 
The direction and the adoption of the new movement, on the other hand, are proposed as role--dependent.
The player with the ball takes a random direction and adopts the movement if the new displacement moves it away from the defender, else it stays on the current position. 
The idea here is to slow down the player movement since it is well--known that the players on ball control, are slower than free players.
The free player, on the other hand, follows a random walk. 
In this regard, our aim is to include in the model the possibility of performing passes of different lengths.
The defender's main role, in turn, is to capture the player with the ball.
Therefore, we consider rule (2.b) as the simplest strategy to choose in the frame of a minimalist model. 

Lastly, the incorporation of the boundaries $R_1$ and $R_2$ is because the development of football games takes place inside confined spaces.
In particular, $R_1$ brings into the model the possibility of capturing short--time ball possession intervals, emulating plays occurring in reduced spaces, as, for instance, fast attacks.
The incorporation of $R_2$, on the other hand, is straightforward, since the real football fields are not limitless. 
The main difference between the real and the model field's bounds is the shape. In this regard, we neglect any possible contribution from the fields' geometry.


We consider that our model offers an adequate balance between simplicity, accuracy, and, as we show in the following sections, empirical validation.
In the Supplementary Material at [URL will be inserted by publisher, S1 and S2], additionally, we show the evaluation of both alternative components and alternative strategies for the model. 
In the following section, we propose a convenient method for tuning the main parameters ruling the model dynamics, (i) the action radius $a$, (ii) the probability of performing a pass $p$, and (iii) the confinement radius $R_1$ and $R_2$.

\subsection{On setting the model's parameters}
\label{se:performance}

The model's performance depends on the correct choice of four parameters: $a$, $p$, $R_1$ and $R_2$.
In this section, we propose a simple method to optimize this tuning procedure.
For the sake of simplicity, we decided to fix $a$, and refer the other radius to this scale, $R_1\to \frac{R_1}{a}$ and $R_2\to \frac{R_2}{a}$.
For the other parameters, we devised a fitting procedure based on the minimization of the sum of the Jensen--Shannon divergences between the observed and the predicted probability distributions of the studied stochastic variables.
To do so, we used the following statistical observables, (i) the distribution of ball possession time $\pt$, 
(ii) the distribution of passes length, $P(\Delta r,Y=\mathrm{Pass})$, and (iii) the distribution of the number of passes performed $\pn$.
With this, we can evaluate the model's dynamics by using three macroscopic variables that we can observe in the real data, a temporal, a combinatorial and a spatial variable describing the interaction between {\it the teammates}.

The method follows the algorithm below, 

\begin{enumerate}
    \item Propose a set of parameters $\rho =(p,R_1,R_2)$;
    
    \item Perform $10^5$ realization, calculate $P(T)$, $P(\Delta r,Y=\mathrm{Pass})$ and $P(N)$ 
    
    \item Compare the three distributions obtained in step 2 with the real data, using the Jensen--Shannon divergence (JSD) \cite{wong1985entropy}. 
    
    \item Propose a new set of parameters $\rho$,
    seeking to lower the sum of the JSD over the three distributions.
    
    \item Back to step 2, and repeat until the JSD is minimized.
\end{enumerate}
Notice, our goal is not to perform a standard non--linear fit but to optimize the search of a realistic set of parameters that simultaneously fit the three distributions.
In this frame, the introduction of the JSD allows us to use a metric distance to compare and assess differences between probability distributions with different physical meanings.
%
%
In the last part of the Supplementary Material at [URL will be inserted by publisher, S1 c.f. FIG.~S4. ], we discuss in detail the implementation of this method.

\section{Results}
\label{se:results}

\subsection{Statistical observables}
\label{se:results1}

\begin{figure*}[t!]
\centering
\includegraphics[width=1\textwidth]{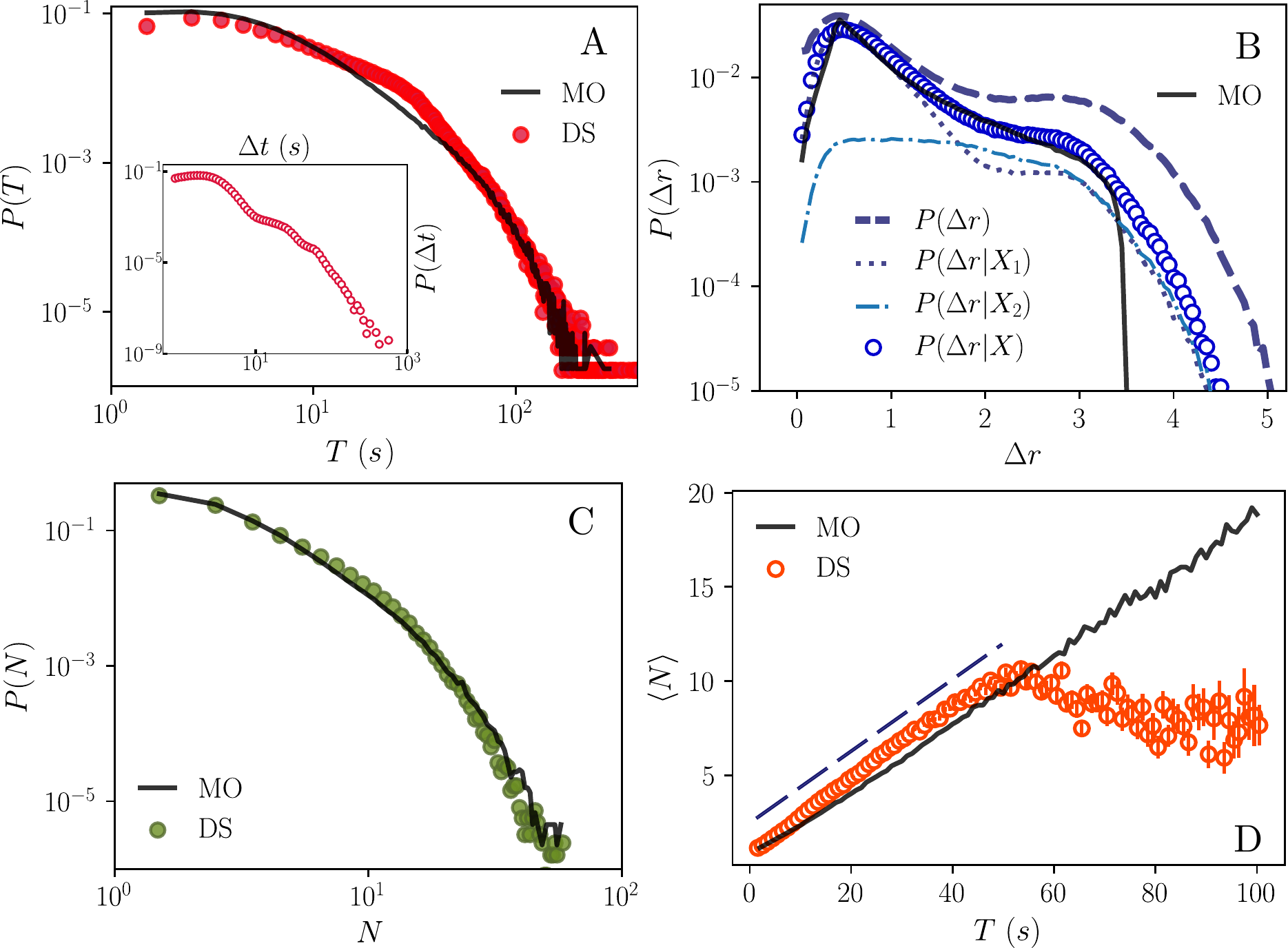}
\caption{ 
Relevant statistical observables found in the dataset {\ttfamily Events} (DS) in ref. \cite{pappalardo2019public}, compared with the model outcomes (MO).
For the results shown in the four panels, we have set the parameters of the model with the values $a=1$, $p=0.3$, $R_1= 2.25$ and $R_2= 16$.
(A) The main plot shows the distribution of the possession time $T$, whereas the inset shows the distribution of the time differences between two consecutive events, $P(\Delta t)$.
(B) Distribution of the distance between two consecutive events segmented in the groups: (i) the whole set of events $P(\Delta r)$, (ii) the passes tagging as sub--type {\it "Simple pass"} $P(\Delta r,X_1)$, (iii) the passes tagging with any other sub--type $P(\Delta r,X_2)$, and (iv) all the passes $P(\Delta r,X)$. Notice, the plot is in linear--log scale.
(C) Distribution of the number of passes in the ball possession intervals, $P(N)$.
(D) Mean value of the number of passes, as a function of the possession time. 
The blue dashed line indicates a linear fit $\avg{N}= \omega_p~ T$ performed on this region, with $\omega_p= 0.19 \pm 0.03$ $(1/s)$.
}
\label{model-exp}
\end{figure*}

The idea of this section is to describe the statistical observables that we extracted from the dataset, and that we use to evaluate the model performance.
The main plot in Fig.~\ref{model-exp} panel A, shows the distribution of possession times. 
We measured the mean value in $\avg{T} = 13.72~s$. 
In this case, we performed a non--linear fit with a function $P(T) \propto T^{-\gamma}$, from where we found $\gamma = 5.1 \pm 0.1$.
We can conclude, despite the distribution seems to follow a power-law behavior, the exponent is large to ensure it \cite{clauset2009power}.
The inset in that panel, in turn, shows the distribution $P(\Delta t)$, the time between two consecutive events.
The same heavy--tailed behavior is observed, which seems to indicate that in both plots, extreme events might not be linked to large values of $T$, but of $\Delta t$. This is probably due to events such as interruptions in the match or similar.
On the other hand, in panel B we show the distribution $P(\Delta r)$, the spatial distance between two consecutive events. 
In this case, we divided the dataset to see the contribution of the event tagged as {\it "Pass"} since, as we show in Fig.~\ref{stats} A, these are the most recurrent entries. 
Let us split $P(\Delta r)$ as follows, 
$P(\Delta r) = 
P(\Delta r, Y=\mathrm{Pass}) + 
P(\Delta r, Y=\mathrm{Other}) $, 
where $Y$ stands for the type of event, the first term is the contribution coming from passes and the second one from any other type of event.   
Moreover, we divided the event pass, into two subtypes 
$P(\Delta r, Y=\mathrm{Pass}) = 
P(\Delta  r, Y=\mathrm{Simple\, pass}) + 
P(\Delta  r, Y=\mathrm{Other\, pass}) $, 
where the first term is the contribution of the sub--type {\it "Simple Pass"} and the second is the contribution of any other sub--type (for example {\it "High pass", "Cross", "Launch"}, etc. c.f. \cite{pappalardo2019public} for further details). 
For the sake of simplicity, hereafter we refer to the type of events {\it "Pass"}, and the subtypes {\it "Simple pass"} and {\it "Other pass"} as $X$, $X_2$ and $X_3$, respectively. 
Notably, we can see a significant contribution of the event {\it "Pass"} to distribution $P(\Delta r)$. 
The peak at $\Delta r \approx 1$ (the mean value) and the hump around $\Delta r \approx 3$ is well explained by the contribution of $P(\Delta r,X)$ and $P(\Delta r,X1)$, whereas $P(\Delta r,X2)$ seems to contributes more to the tail. 
This multi-modal behaviour, likewise, might evidence the presence of two preferential distances, from where teammates are more likely to interact by performing passes.
Panel C shows the distribution $P(N)$ of the number of passes per BPI. We observe the presence of a heavy tail at the right.
The mean value, $\avg{N}= 3.1$, indicates that on average we observe $\approx 3$ passes per BPI.
Concerning this point, in panel D, we show the relation between the number of passes and the possession time. Interestingly, we observe a linear relation for values within $0<T< 60~(s)$ (see solid blue line in the panel). 
From our best linear fit in this region, we obtain $\avg{N}(T)= \omega_p~T$ with $\omega_p=0.19 \pm 0.03$ ($R^2=0.99$).
This parameter can be thought in overall terms as the rate of passes per unit of time. Therefore, we conclude that during ball possession intervals, $\approx 0.2$ passes per second are performed.

\subsection{Assessing the model performance}

In this section, we evaluate and discuss the model's outcomes.
The results are shown in Fig.~\ref{model-exp}. Panels A, B, C, and D show the comparison between the results obtained from the  dataset (discussed above) and from the model's simulations (black solid lines).
We used the set of parameters $(p,a,R_1,R_2) = (0.3,~1,~2.25,~16)$. 

For the distribution $P(T)$ in panel A, we obtain a Jensen-Shannon distance of $D_{JS}= 0.017$, which indicates a good similarity between the dataset and the model results.
However, we observe a shift in the mean of $\approx -20\%$, and a problem to capture "the hump" of the curve  around $T \approx 30~s$.
For the distribution of passes length, $P(\Delta r,X)$, showed in panel B, we observe a very good similarity $D_{JS}= 0.008$. Moreover, we can see the model succeeds in capturing the bimodality of the distribution, which seems to indicate that the proposed model rules are very effective for capture both nearby and distant passes, two interaction distances. 
On the other hand, the model fails in capturing the tail, possibly because these events are related to very long passes ({\it "Goal kicks"} or {\it "Cross passes"}) not generated by the simple dynamics of the model.
In panel C, we show the distribution of the number of passes $P(N)$. 
The calculation for the Jensen--Shannon distance gives the value $D_{JS}=0.0007$, which indicates a very good similarity between the curves. 
In this case, the value of $p$ seems to be crucial. 
Note as the chosen value for $p$ is near to the rate $\omega_p = 0.19$ passes per second, reported in the previous section.
Regarding the relation $\avg{N}$ vs. $T$ in panel D, the dataset shows that, on average, the number of passes cannot indefinitely grow with the possession time, which is likely a finite--size effect. 
Our simple model, in turn, allows the unrealistic unbounded growth of $\avg{N}$.

Lastly, let us put the parameter values in the context of real football dimensions. 
Regarding the action radius $a$, the literature includes reported estimations from kinetic and coordination variables~\cite{schollhorn2003coordination, lames2010oscillations}, where speed measurements~\cite{little2005specificity, loturco2019maximum}
show that professional players are able to move in a wide range within $1.1$ -- $4.8$ $m/s$. 
Thus, it would be easy for a professional player to control a radius of $a \approx 2~m$.
If we set this value for $a$, we proportionally obtain for the internal and the external radius, the values  $R_1\approx5~m$ and $R_2\approx 32~m$, respectively. 
Consequently, in the frame of our model, the dynamics of the possession intervals take place into areas within a range of $78~m^2$  (approximately a goal area), and $3200~m^2$ ($\approx 47 \%$ of the Wimbledon Greyhound Stadium).
Therefore, we conclude the proposed parameters are in the order of magnitude of real football field dimension, and we can confirm that the dynamics of the model is ruled upon a realistic set of parameters' values. 

\subsection{Mapping the model in a theoretical framework}
\label{se:results2}

\begin{figure*}[t!]
\centering
\includegraphics[width=1\textwidth]{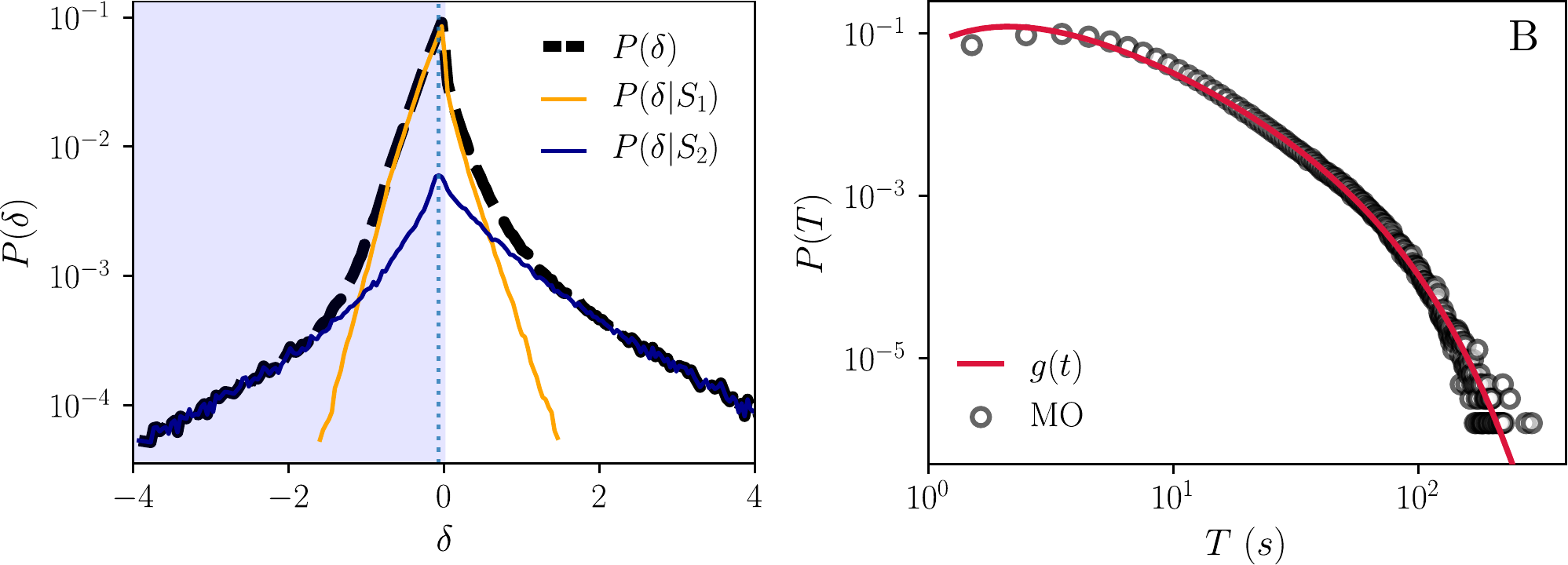}
\caption{Results of mapping the model to a Wiener process with drift and an absorbing barrier.
(A) distribution of steps $\delta$, segmented in all the data, $P(\delta)$, those steps given in the context of a simple persecution, $P(\delta,S_1)$, and those steps in the context of a pass, $P(\delta,S_2)$ 
(B) non--linear fit performed to distribution $P(T)$ (MO), using the expression $g(t)$ given by eq.~(\ref{eq:mpt}).
}
\label{fi:gt}
\end{figure*}

We propose a theoretical framework to understand the distribution of possession times, $P(T)$, observed from the model's outcomes. 
Every realization can be thought of as a process where the defender must capture the ball. A ball that, due to the movements and passes performed by the teammates, may follow a complicated path in the plane. However, since the defender always takes the direction towards the ball, the process can be reduced to a series of movements in one dimension.
To visualize this mapping we fix the origin of our 1D coordinate system at the ball position and define the coordinate $x$ of the defender as the radial distance $d$ between the ball and the defender. In this frame, the defender takes steps back and forth depending on whether the radial distance between the ball and defender is increasing or decreasing, respectively.
The step size $\Delta d$ of this random walk is variable, and the process ends when the coordinate $x$ of the defender reaches the interval $(-a, a) $ (c.f. Section II.~C, rule 8).  
In this process, the step size distribution characterizes the random walk. 
Let us define $\delta = \Delta d/d_0$ as the step size normalized to the initial distance between the players.  
Then, in Fig.~\ref{fi:gt} A, we plot the distribution $P(\delta)$ analyzing two possible contributions for the steps, (i) the steps taken when the defender follows the player with the ball ($S_1$), (ii) those generated when a pass between teammates occurs ($S_2$). In order to visualize these contributions, we have plotted $P(\delta)$, and the joint probabilities $P(\delta,S_1)$ and $P(\delta,S_2)$, fulfilling $P(\delta)= P(\delta,S_1)+P(\delta,S_2)$. 
From this perspective, we can see that $(S_2)$ explains the extreme events whereas $(S_1)$  explain the peak

On the other hand, if we measure the mean value of both contributions we obtain $\avg{\delta}_{P(\delta,S_1)} = -0.14$, $\avg{\delta}_{P(\delta,S_2)} = 0.22$, which means that in average, the first contribution brings the defender towards the ball, and the second takes it away. However, notice that the full contribution is negative, $\avg{\delta}_{P(\delta)} = -0.07$, which indicates the presence of a drift leading the defender towards the ball. 

From this perspective, we can map the dynamics to a random walk with drift, and in the presence of an absorbing barrier.
Moreover, in the approximation where $\delta$ is constant, the process described above is governed by the following Focker--Plank equation,
\begin{equation}
    \frac{\sigma^2}{2} \frac{\partial^2 p}{\partial x^2}- \mu \frac{\partial p}{\partial x} = \frac{\partial p}{\partial t} 
\label{eq:f-p}
\end{equation}
subject to the boundary conditions,
\begin{align*}
    p(d_0,x;0) &= \delta (x), \\
    p(d_0,x_b;t) &= 0,
\end{align*}
where $p(d_0,x,t)$ is the probability of finding a walker that starts in $d_0$, in the position $x$ at time $t$.
The coefficients $\mu$ and $\sigma$ are the drift and the diffusion, and $x_b$ indicates the position where the absorbing barrier is placed. Additionally, it can be proved that the probability distribution of the first passage time $\tau$, for a walker reaching the barrier, is given by \cite{cox1977theory},
\begin{equation}
    g(\tau) =  \frac{x_b}{\sigma \sqrt{2\pi \tau^3}} 
    \exp \left(-\frac{(x_b -\mu \tau)^2}{2 \sigma^2 \tau} \right),
\label{eq:mpt}
\end{equation}
which can be straightforwardly linked to the distribution of possession times $P(T)$.

In this theoretical framework, we used eq. (\ref{eq:mpt}) to perform a non--linear fit of $P(T)$, via the parameters $\mu$ and $\sigma$. 
We set $x_b= a$, as the action radius can be thought of as the barrier's position. 
The result presented in Fig.~\ref{fi:gt} B, shows the fitting is statistically  significant, yielding a correlation coefficient $r^2 = 0.97$, with $\mu= 0.09 \pm 0.02$ and  $\sigma= 0.39 \pm 0.03$. Moreover, notice that we achieve a very good agreement between the drift value and $\avg{\delta}_{P(\delta)}$, in magnitude.
Therefore, we can conclude that, in the context of the model, a random walk with a constant step $\delta$ and a drift $\mu$, is a good approximation for a walker drawing steps from $\avg{\delta}_{P(\delta)}$. Furthermore, this approximation explains the long tail observed in $P(T)$ for both, the outcomes of the model and the empirical observations. 


\section{Discussion}

In this contribution, we focused on analyzing the dynamic of ball possession intervals. 
We have performed an empirical study of a novel dataset, detected relevant statistical patterns, and on this base, proposed a numerical agent--based model.
This model is simple, and it can be easily interpreted in terms of the features of the phenomenon under discussion.
Moreover, we proposed a theoretical interpretation of the numerical model in the frame of an even simpler but better--understood physical model: the Wiener process with drift and an absorbing barrier. 
In this section, we extend the discussion regarding these results.

First, we fully characterize BPIs of the extensive dataset that compiles most of the events during the games, identifying the main contributions.
Four salient features were identified and used later as the input to devise a minimalist football model to study the dynamics of ball possession intervals.
Namely, (i) the most frequent type of events, (ii) events leading to a change in possession, (iii) the number of players participating in a BPI, and (iv) the different types of events during BPIs.
We found that the most frequent event is {\it "Pass"}, which twice the second most common event, {\it "Duels"}. The latter, in turn, is the most common type of event triggering ball possession changes. 
In most cases, just two players are involved in a BPI, and during a BPI there are usually at most two events. 

Prompted from these findings, we introduced a minimalist model composed of two {\it teammates} and a single {\it defender}, that following simple motion rules, emulates both on--ball and off--ball actions. 
This model can be tuned by setting four independent parameters 
$a$, $p$, $R_1$ and $R_2$, which control the action radius, the probability of making a pass, and the internal and external radius, respectively.

We evaluated the model's performance by comparing the outcomes with three statistical observables in the possession intervals, the distribution of possession time $\pt$, the distribution of passes length $\pr$, and the distribution of the number of passes $\pn$.
To this end, we have introduced a simple method based on the evaluation of the Jensen--Shannon distances, as a criterion to fit the simulation's outcomes to the real data.
Remarkably, despite the simplicity of the model, it approaches very well the empirical distributions. 

Finally, to get a physical insight into the process behind ball possession dynamics, we map the model to a one--dimensional random walk in which the ball is fixed at the origin, and the defender moves taking non--uniform steps of length $\delta$.
We showed that since $\avg{\delta}_{P(\delta)}<0$ holds, the defender moves following a preferential direction towards the ball. 
Then, we can use the theoretical framework of a Wiener process with drift and an absorbing barrier to describe the model's dynamics. 
We evaluated this hypothesis by performing an non--linear fit to the distribution of possession times, $P(T)$, with the expression of the first passage time for the Wiener process, finding a very good agreement.
The mapping shows that the agents' dynamics in the numerical model can be understood in the frame of a simple physical system. 

We can think in the game of football as a complex system where the interactions are based on cooperation and competition.
Competition is related to teams' strategies, it concerns the problem of how to deal with the strengths and weaknesses of the opponent \cite{hewitt2016game}. 
Strategies are usually previously planned and are developed during the entire game, hence it could be associated with long--term patterns in the match.
Cooperation, on the other hand, can be linked to tactical aspects into the game. 
Where interactions bounded to a reduced space in the field, short periods into the match, and carry out by a reduced number of players,
could be associated with short--term patterns.
Ball possession intervals are related to cooperative interactions.  
Therefore, in this work, we are not studying the full dynamics of a football match but tactical aspects of the game.
In this frame, our work should be considered as a new step towards a better understanding of the interplay between the short-term dynamics and the emerging long-term patterns within the game of football when studied as complex systems with non-trivial interaction dynamics.

From a technical point of view, our model could be used as a starting point to simulate and analyze several tactical aspects of the game. Note that the main advantage of our simple numerical model is that it easily allows the introduction of complexity: more players, different types of interactions, etc.
For instance, simulations based on our model can be useful to design training sessions of small--sided games \cite{sangnier2019planning,eniseler2017high,reilly2005small}.
Where coaches expose players to workout under specific constraints: in reduced space, with a reduced number of players, with coordinated actions guided by different rules, etc \cite{sarmento2018small}.
Moreover, by performing simulations is possible to estimate the physical demand of the players, which is useful for sessions planning and post evaluation \cite{hodgson2014time}.

Lastly, as we said above, we consider that a full characterization of the football dynamics should focus on the study of both competitive and cooperative interactions.
In this work, we focus on the latter, a first--step to address the former could focus on analyzing the spatiotemporal correlations between consecutive possession intervals.
In this regard, we let the door open to futures research works in the area.


\begin{acknowledgments}
This work was partially supported by grants from CONICET (PIP 112 20150 10028), FonCyT (PICT-2017-0973), SeCyT–UNC (Argentina) and MinCyT Córdoba (PID PGC 2018). 

\end{acknowledgments}

\clearpage
\newpage

\end{document}